# LOW POWER LOW VOLTAGE BULK DRIVEN BALANCED OTA

Neha Gupta[1], Sapna Singh[2], Meenakshi Suthar[3] and Priyanka Soni[4]

Faculty of Engineering Technology, Mody Institute of Technology and Science, Lakshmangarh, Sikar, India
[1]wdnehagupta@gmail.com, [2]singh.sapna067@gmail.com [3]meenakshi.suthar32@gmail.com,
[4]priyankamec@gmail.com

*ABSTRACT*

*The last few decades, a great deal of attention has been paid to low-voltage (LV) low-power (LP) integrated circuits design since the power consumption has become a critical issue. Among many techniques used for the design of LV LP analog circuits, the Bulk-driven principle offers a promising route towards this design for many aspects mainly the simplicity and using the conventional MOS technology to implement these designs. This paper is devoted to the Bulk-driven (BD) principle and utilizing this principle to design LV LP building block of Operational Transconductance Amplifier (OTA) in standard CMOS processes and supply voltage 0.9V. The simulation results have been carried out by the Spice simulator using the 130nm CMOS technology from TSMC.*

## 1. INTRODUCTION

The modern electronics industry is witnessing the dominance of miniaturization in every sphere of electronics and communications forming the backbone of medical electronics, mobile communications, computers, state-of-art processors etc. Designing high performance analog circuits is becoming increasingly challenging with the persistent trend towards reduced supply voltages. An important factor concerning analog circuits is that, the threshold voltages of future standard CMOS technologies are not expected to decrease much below what is available today. Though the MOS transistor is a four terminal device, it is most often used as a three terminal device since the bulk terminal is tied either to the source terminal. Therefore, a large number of possible MOS circuits are overlooked; hence a good solution to overcome the threshold voltage is to use the Bulk-driven principle [1-3].

The principle of the Bulk-driven method is that, the gate-source voltage is set to a value sufficient to form an inversion layer, and an input signal is applied to the bulk terminal. Since the bulk voltage affects the thickness of the depletion region associated with the inversion layer i.e. the conduction channel. The drain current can be modulated by varying the bulk voltage through the body [4-6]. In the present work a Low Power Low Voltage Bulk driven Balanced Operational Transconductance Amplifier has been designed. In this design inputs are applied to the bulk terminal of the NMOS input transistors and gate terminals are biased with a voltage so that the NMOS transistors come in active region. In this manner, the threshold voltage can be either reduced or removed from the signal path.

## 2. BULK- DRIVEN BALANCED OTA

When the input devices of the bulk-driven differential pair in Figure 1. operate in the strong inversion saturated region, their drain current is given by:

$$I_D = \frac{1}{2} k'_p \left(\frac{W}{L}\right)(V_{SG} - |V_{TH}|)^2 \qquad (1)$$

Where, the symbols have their usual meaning and the channel length modulation effect has been neglected. The behavior of the threshold voltage as a function of the bulk-to-source voltage VBS of the input devices can be expressed as

$$|V_{TH}| = |V_{TH0}| + |\gamma|\left[\sqrt{2|\Phi_F| + V_{BS}} - \sqrt{2|\Phi_F|}\right] \qquad (2)$$

Where, $V_{TH0}$ is the value of the threshold voltage $V_{TH}$ when $V_{BS}$ is zero, $\gamma$ is the body effect parameter, and $\Phi F$ is Fermi potential. The operation of bulk-driven devices is based on the body effect, that is, on the dependence of $V_{TH}$ on $V_{BS}$ [7-8]. Indeed, from (1) and (2), $I_D$ changes when changing $V_{BS}$ and, hence, a transconductance function between the bulk voltage and the drain current is achieved. The transconductance of a bulk-driven MOS transistor may be obtained from (1) and (2) as

$$g_{mb} = \frac{\partial I_D}{\partial V_{SB}} = \frac{|\gamma|g_m}{2\sqrt{2|\Phi_F| + V_{BS}}} = \frac{|\gamma|}{2\sqrt{2|\Phi_F| + V_{BS}}}\sqrt{2\beta I_D} \qquad (3)$$

Where, $g_m$ is the gate transconductance of the device. If typical values for $\gamma$ and $\Phi F$ are considered, the value of $g_{mb}$ ranges from 20% to 50% the value of $g_m$.

To enable body driving, one must first bias the gate to develop a channel inversion layer. Once the inversion layer has been formed there will be depletion region associated with junction of the inversion layer and transistor body. The current is then modulated by varying the body voltage, which in turn modulates the inversion layer width via the depletion region. The drain current versus body voltage for this condition is obtained from (1) and (2)

$$I_D = \frac{1}{2} k'_p \left(\frac{W}{L}\right)\left(V_{GS} - V_{T0} - \gamma\sqrt{2|\Phi_F| + V_{BS}} + \gamma\sqrt{2|\Phi_F|}\right)^2 \qquad (4)$$

## 3. BULK DRIVEN BALANCED OTA

The Bulk-driven OTA was presented in [9-12]. The two-stage OTA is shown in Figure1. It consists of two stages, the first which is combined of the Bulk-driven differential stage with NMOS input devices M1 and M2 and the current mirror M3 and M4 acting as an active load. By setting the gate-source voltage to a value sufficient to turn on the transistor, then the operation of the Bulk-driven MOS transistor becomes a depletion type. Input voltage is applied to the bulk-terminal of the transistor to modulate the current flow through the transistor. OTA with Bulk-driven input transistors has been designed. The design is depicted in Figure 1 and the simulation results are shown in Table 1.

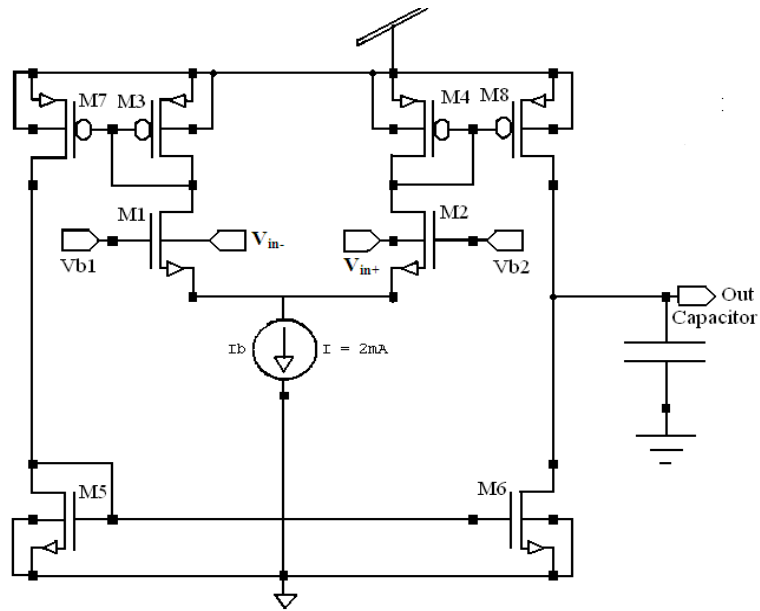

Figure 1 Bulk Driven Balanced OTA

## 4. BULK DRIVEN BALANCED OTA CHARACTERISTICS

Operating the MOS through the bulk-terminal allows the design of extremely low-supply voltage circuits. The behavior of the bulk-driven MOSFET is very close to a junction-field-effect transistor (JFET), the signal is applied between the bulk and the source and the current flowing from the source to the drain is modulated by the reverse bias applied on the bulk. The gate-source voltage of the MOS is fixed and is set to turn the MOSFET on. Desirable characteristics of Bulk-driven transistors are [10, 11]:

- Low-voltage low-power consumption of the circuits.
- Design simplicity and the acceptable features of the circuits.
- Depletion characteristics avoid VT requirement in the signal path.
- The conventional front gate could be used to modulate the Bulk-driven MOS transistor.

Undesirable characteristics of Bulk-driven transistors are:

- The transconductance of a Bulk-driven MOST is smaller than that of a conventional Gate-driven, which may result in lower GBW in OTA.
- The polarity of the Bulk-driven MOST is technology related. For a P (N) well CMOS process, only N (P) channels Bulk-driven MOSTs are available. This may limit its applications. For example a rail-to-rail Bulk-driven op-amp needs a dual well process to realize it, this process is more expensive, bigger chip area needed and it has worst matching comparing with one well process.
- Prone to turn on the bulk-channel PN junction, which may result in a latch-up problem.

### 4.1 AC Response

This response is used for observing open loop gain, Unity Gain Bandwidth (UGB), 3-dB bandwidth and the Phase Margin of the circuit. In the test setup a differential AC signal of 1V is applied to the inputs. No dc bias potential is applied to bulk terminals. The bias voltage is applied to gate terminals to bring transistor in saturation region. The output was taken between single end and ground terminal.

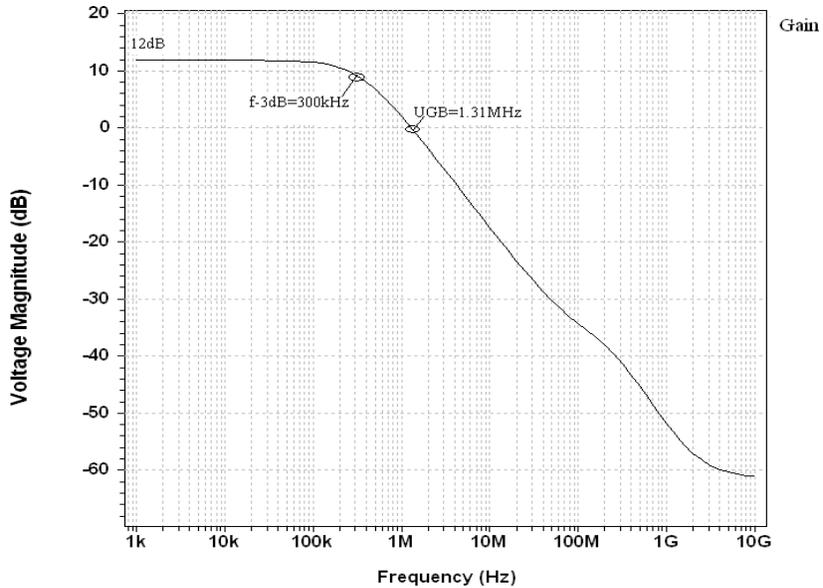

Figure 2. Voltage Gain of Bulk Driven Balanced OTA

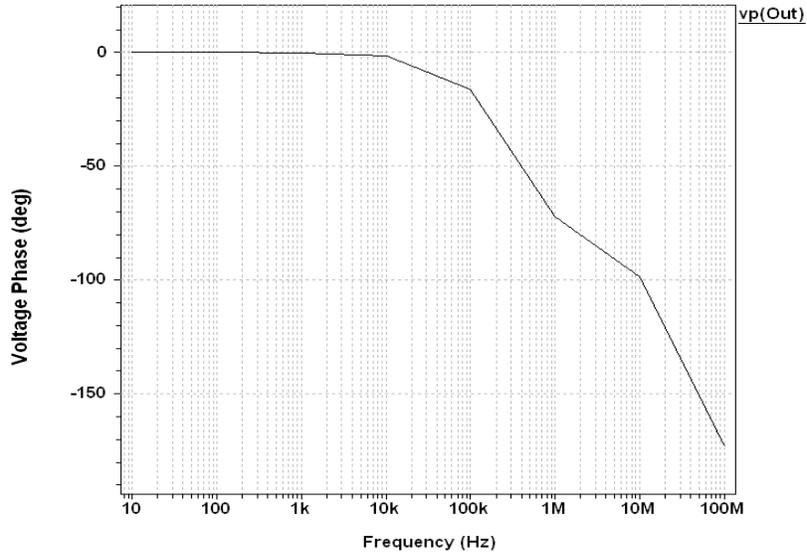

Figure 3. Voltage Phase of Bulk Driven Balanced OTA

The Gain of Bulk Driven Balanced OTA is 12 dB , 3dB bandwidth is 300KHz, Unity Gain Bandwidth is 1.31MHz, Phase Margin is (-73º+180º) 107º

## 4.2 Step Response

The slew rate is defined as the ratio of the maximum output current to the load capacitance as shown below:

$$SR = \frac{dV_{out}}{dt} = \frac{I_{out}}{C_L} \tag{5}$$

It is specified on the amplifier's data sheet in units of V/µs. It implies that if the input signal applied to an amplifier circuit is of the nature that demands an output response faster than the specified value of SR, the operational amplifier will not comply. The settling time of an amplifier is very important parameter as it defines the speed at which amplifier can be sampled.

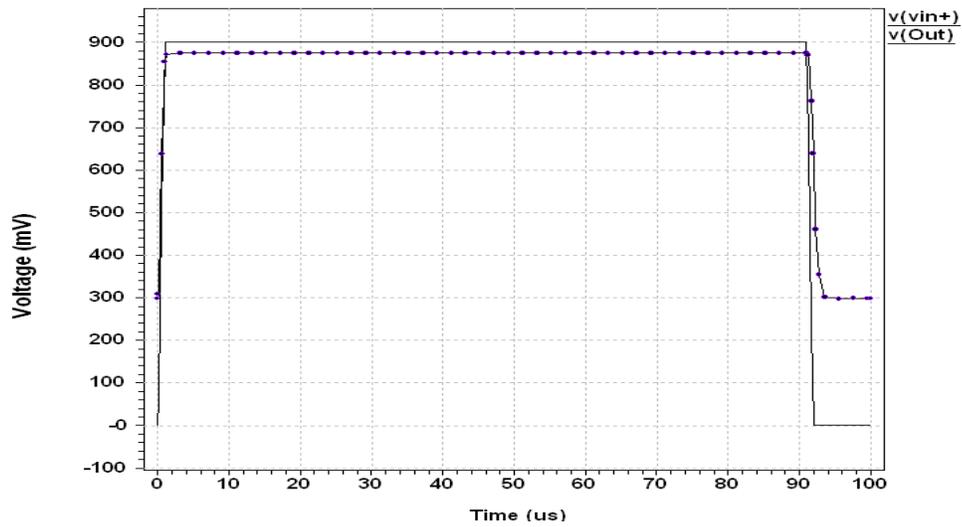

Figure 4. Slew rate of Bulk Driven Balanced OTA

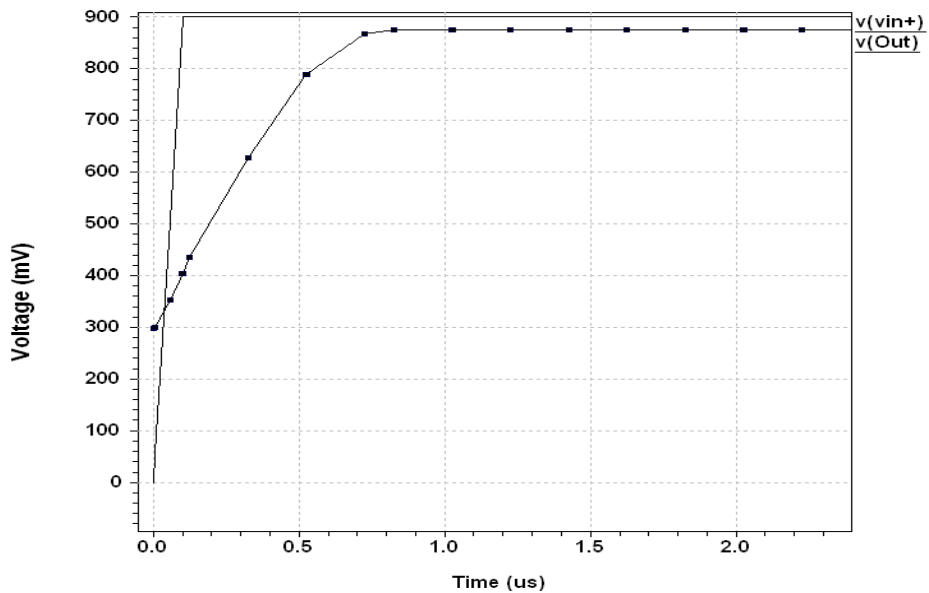

Figure 5. Settling Time of Bulk Driven Balanced OTA

For measurement of slew rate a step signal of 0.9V is applied at non inverting terminal of the amplifier. The amplifier's slew rate is 543mV/µs for rising edge and 226mV/ µs for the falling edge and settling time of is 0.7µs

### 4.3 Effect of Variation of Temperature on Frequency response

The frequency response is shown in Figure 6. at different temperatures. The temperature is varied from -20º to 70 º and corresponding change in gain is from 11.28dB to 12.43dB. The gain will decrease with temperature because transconductance of opamp will decrease with temperature. The decrement in transconductance is due to decrement in mobility.

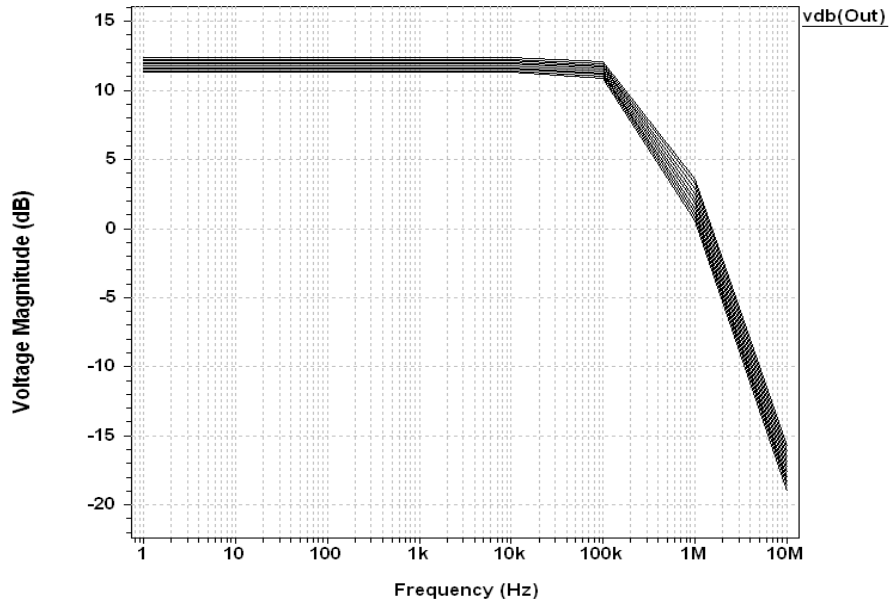

Figure 6. Effect of Variation of Temperature on Frequency response

### 4.4 Transient Results

We get the output swing (peak to peak) 550mV. In this we can apply an input signal of swing 400mV for which the output signal swing is not distorted. Simulation result is shown Figure 7.

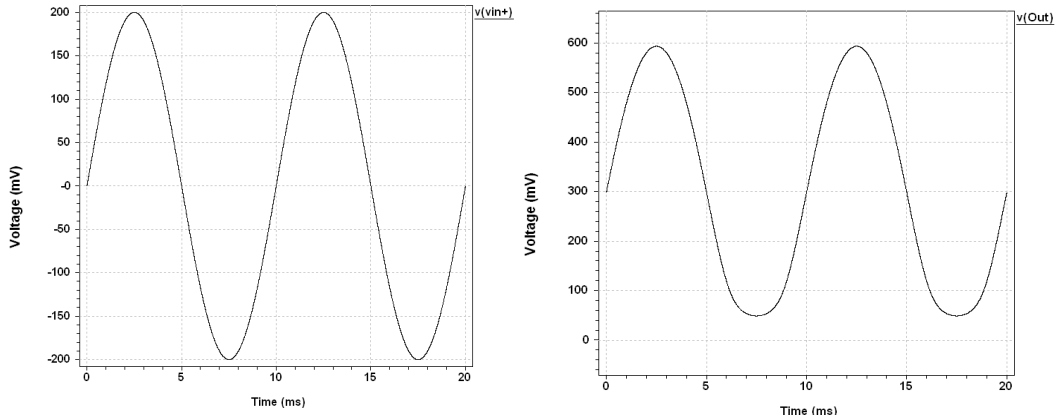

Figure 7. Output Voltage Swing of Bulk Driven Balanced OTA

### 4.5 Power Measurement

Average power consumed at 0.9 V supply is 3.9µW. Power consumption at different power supplies are shown in Figure 8.

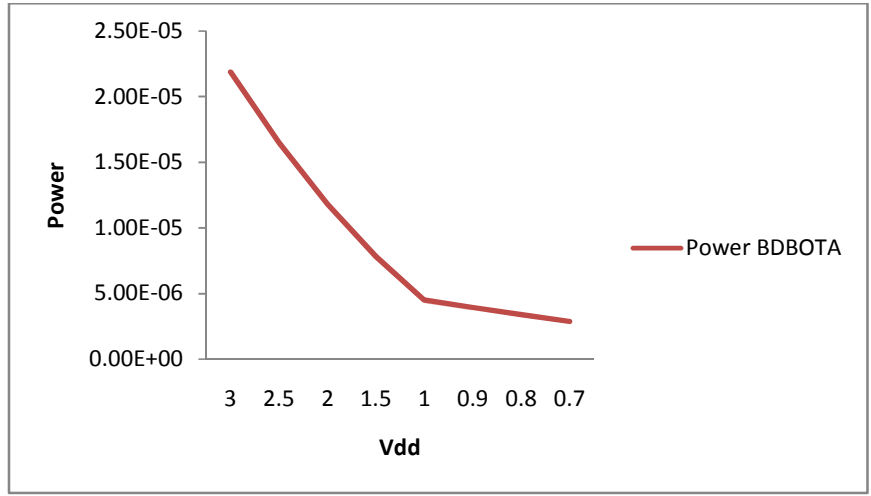

Figure 8. Power measurements with different supply voltages (Vdd)

### 4.6 Noise Analysis

The input referred noise is 400nV/$\sqrt{Hz}$ and output referred noise is 800 nV/$\sqrt{Hz}$. The simulations for noise analysis are shown in Figure 9(a) and 9(b).

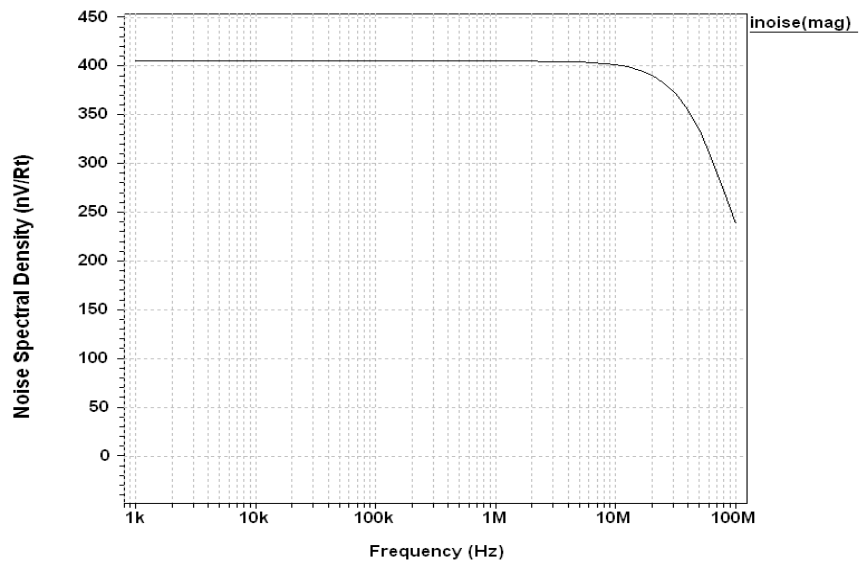

Figure 9(a). Input referred noise

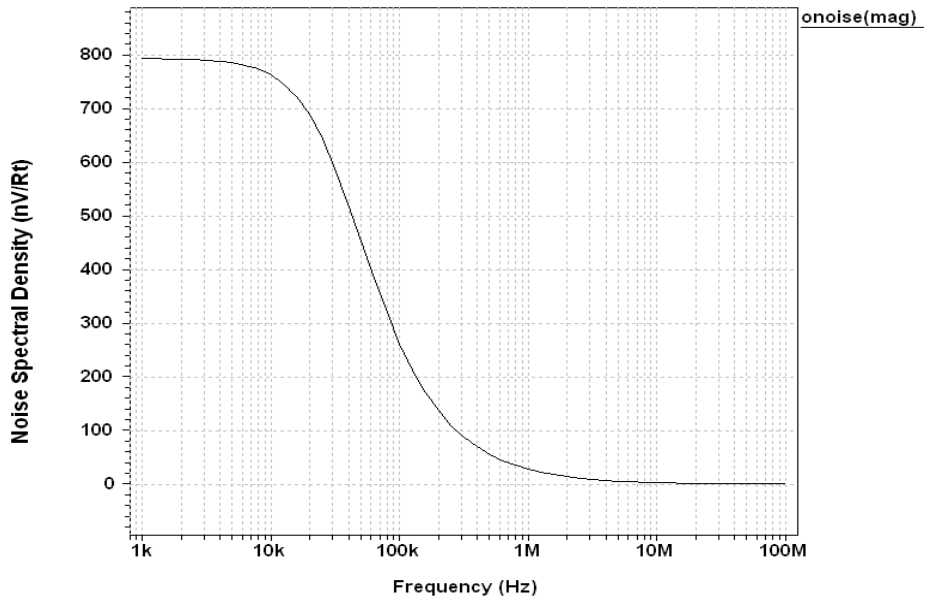

Figure 9(b). Output referred noise

Prepared circuit of amplifier is simulated at Tanner EDA T-SPICE tool at 130nm technology. The amplifier is to be powered from a 0.9 volts power supply. All values have been measured at load capacitance of 1 pF. The constant current source $I_b$ is 20μA.

### 4.7 DC OTA Response at Various Common Mode Voltages

The output voltage is measured at different differential input (VVdiff) ranging from -0.5V to 0.9 V .This simulation is done at various common mode voltages(Vcm) ranging from 0 to 0.5V. The simulation is shown in Figure 10.

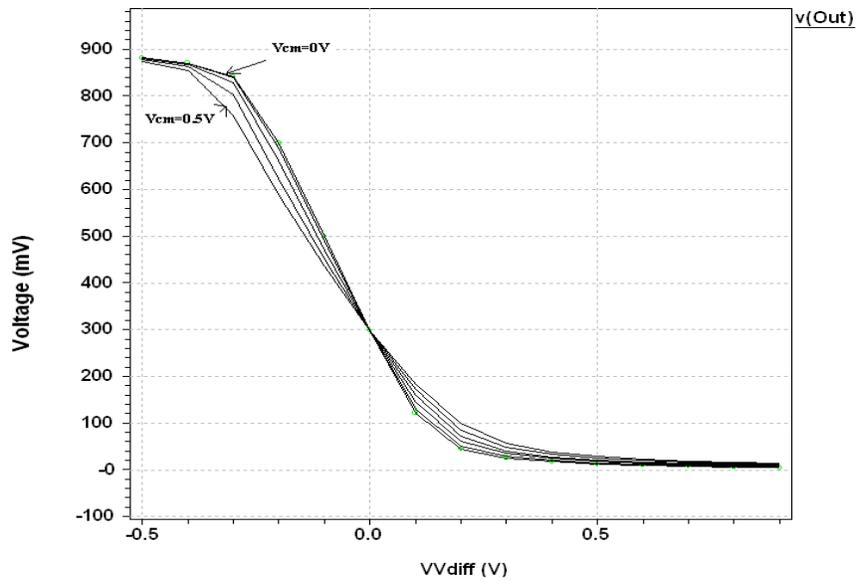

Figure 10. Measured DC OTA responses at different common mode input voltage

Also output voltage is plotted at differential input ranging from 0 to 900mV in Figure 11. In this graph we can see that output is linearly following the input over a range of input voltage from 200 to 800mV. The linear range is 600mV.

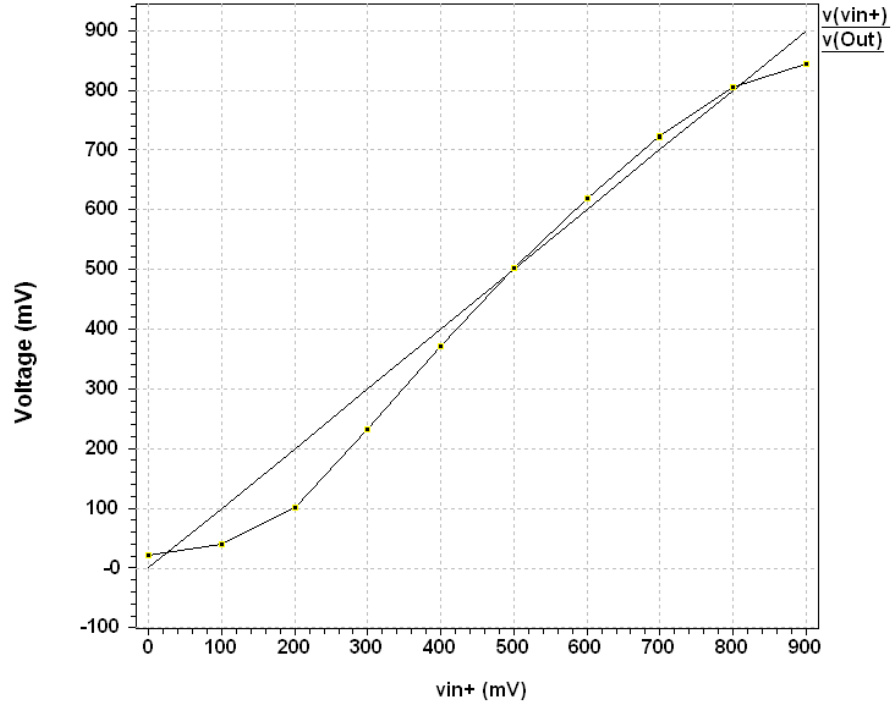

Figure 11. Output Voltage is plotted against dc sweep of input voltages

Table 1 Simulated Results of Bulk Driven Balanced OTA

| Characteristics | Simulated results |
|---|---|
| Power consumption | 3.9μW |
| Open loop gain | 12 dB |
| Phase Margin | 107º |
| 3 dB Bandwidth | 300kHz |
| Unity Gain Bandwidth | 1.31MHz |
| Positive Slew Rate | 543mV/μs |
| Negative Slew Rate | 226mV/μs |
| Settling Time (at 10% tolerance) | 0.7μs |
| Maximum voltage swing | 550mV |
| Output Referred Noise | 800nV/$\sqrt{Hz}$ |
| Input Referred Noise | 400nV/$\sqrt{Hz}$ |

## 5. CONCLUSION

In this paper a two stage low power low voltage Bulk Driven Balanced OTA has been designed and simulated. In this circuit as inputs are applied on bulk terminal so threshold voltage limitation is not there, i.e. we can have input of more voltage swing to obtain output swing without distortion. The output voltage swing is 550 mV and input referred noise of the proposed bulk driven circuit is 400 nV/$\sqrt{Hz}$ and output referred noise is 800 nV/$\sqrt{Hz}$. This circuit is operated at a power supply of 0.9V and the power dissipation is 3.9μW.

As power and voltage are the two main constraints in analog/mixed signal systems. Bulk driven NMOS input transistors are used for low power and low voltage design. The ac signal is applied to Bulk terminal and gate terminal is at ac ground. The current in NMOS transistor is modulated by bulk signal which is a novel technique to avoid threshold voltage limitation in the path of input ac signal. Gate terminal is only biased to bring the transistor in active region. This technique uses all the four terminals of MOS device as bulk terminal is ignored in earlier years and is not fully utilized to improve device performance.

## 6. ACKNOWLEDGEMENT

We would like to sincerely thank Prof. B.P Singh, Head of Department of Electronics & Communication, Mody Institute of Technology and Science, Lakshmangarh who inspired us to do this work. In addition we would like to thank Prof P.K Das, Dean, Faculty of Engineering, Mody Institute of Technology and Science for providing us resources to carry out our work.

**AUTHORS**

Neha Gupta received her B.Tech degree in Electronics and Communication Engineering from Mody institute of Technology and Scien ce, Lakshmangarh in 2009, and is currentlyM.Tech VLSI Design student of Mody Institute of Technology and Science, Lakshmangarh. Her research interests are analog integrated circuit design and high performance analog circuits

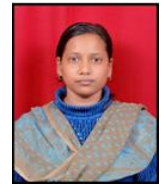

Sapna Singh received her B.Tech degree in Electronics and Communication Engineering from Kumaon Engineering College, Dwarahat in 2010,and is currently M.Tech VLSI Design student of Mody Institute of Technology and Science, Lakshmangarh. Her research interests are low power vlsi design and analog and digital integrated circuit design.

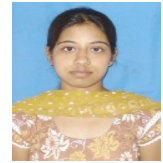

Meenakshi Suthar received her B.Tech degree in Electronics and Communication Engineering from Marudhar Engineering College,Bikaner in 2010,and is currently M.Tech VLSI Design student of Mody Institute of Technology and Science, Lakshmangarh. Her research interests are analog integrated circuit design and analog communication systems.

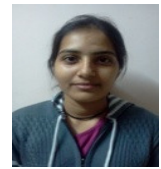

Priyanka Soni received her B.Tech degree in 2007 from Marudhar Engineering College in Bikaner and M.Tech degree in VLSI design in 2011 from Mody Institute of Technology and Science, Lakshmangarh . Currently she is working as Assistant Professor in Mody Institute of Technology and Science .Her research interest are analog integrated circuit design and low power vlsi design.

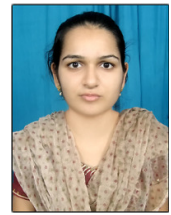